\documentclass[journal=jpclcd,manuscript=letter,layout=twocolumn]{achemso}

\usepackage[version=3]{mhchem} 
\usepackage{amssymb} 
\usepackage{hyperref}
\usepackage{color}
\usepackage{pdfpages} 

\author{Michele Invernizzi}
\affiliation{Department of Mathematics and Computer Science, Freie Universit\"at Berlin, 14195 Berlin, Germany}
\email{michele.invernizzi@fu-berlin.de}

\author{Andreas Kr\"amer}
\affiliation{Department of Mathematics and Computer Science, Freie Universit\"at Berlin, 14195 Berlin, Germany}

\author{\\Cecilia Clementi}
\affiliation{Department of Physics,  Freie Universit\"at Berlin, 14195 Berlin, Germany}
\alsoaffiliation{Department of Chemistry, Rice University, 77005 Houston, United States}
\alsoaffiliation{Center for Theoretical Biological Physics, Rice University, 77005 Houston, United States}

\author{Frank No\'e}
\affiliation{Microsoft Research AI4Science, 10178 Berlin, Germany}
\alsoaffiliation{Department of Mathematics and Computer Science, Freie Universit\"at Berlin, 14195 Berlin, Germany}
\alsoaffiliation{Department of Physics,  Freie Universit\"at Berlin, 14195 Berlin, Germany}
\alsoaffiliation{Department of Chemistry, Rice University, 77005 Houston, United States}
\email{franknoe@microsoft.com}

\title{Skipping the Replica Exchange Ladder \\with Normalizing Flows}

\abbreviations{MD, REX, NF, LFEP, LREX, FES}

\begin{document}

\newlength{\myfigwidth}
\setlength{\myfigwidth}{\columnwidth} 

\begin{tocentry}




%
\includegraphics[width=\columnwidth]{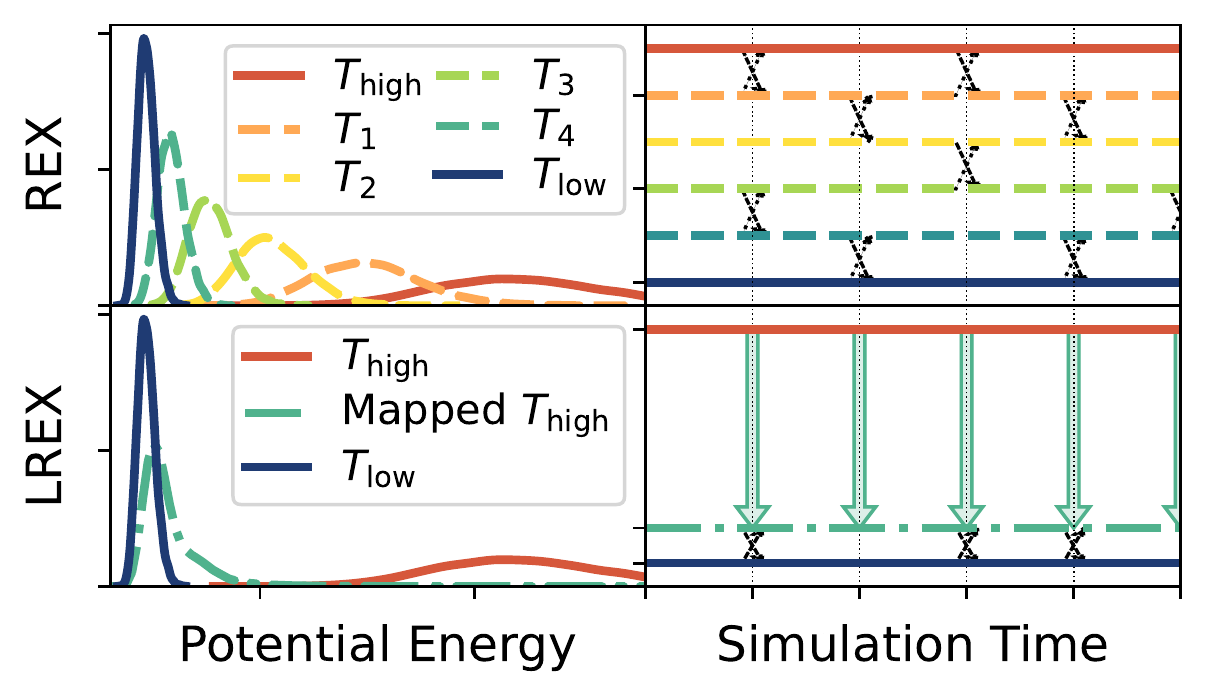}
\end{tocentry}

\begin{abstract}
We combine replica exchange (parallel tempering) with normalizing flows, a class of deep generative models.
These two sampling strategies complement each other, resulting in an efficient strategy for sampling molecular systems characterized by rare events, which we call learned replica exchange (LREX).
In LREX, a normalizing flow is trained to map the configurations of the fastest-mixing replica into configurations belonging to the target distribution, allowing direct exchanges between the two without the need to simulate intermediate replicas.
This can significantly reduce the computational cost compared to standard replica exchange.
The proposed method also offers several advantages with respect to Boltzmann generators that directly use normalizing flows to sample the target distribution.
We apply LREX to some prototypical molecular dynamics systems, highlighting the improvements over previous methods.
\end{abstract}

Molecular simulations are becoming more and more important for studying complex phenomena in physics, chemistry, and biology.
One of the long-lasting challenges for molecular simulations is to efficiently sample the equilibrium Boltzmann distribution, which is often characterized by multiple metastable states.
In this letter, we propose a novel sampling method that combines the well-established replica exchange method\cite{Swendsen1986,Earl2005} with a recent machine-learning technique known as normalizing flows\cite{Tabak2010,Tabak2013,Rezende2015}.
To this end, we first summarize the two original methods and discuss their strengths and weaknesses.
We then introduce flow-based exchange moves and show with a few prototypical examples how they help outperform classical replica exchange.

Replica exchange (REX), also known as parallel tempering, is a popular enhanced sampling method\cite{Swendsen1986,Earl2005}.
It uses a ladder of overlapping distributions to connect the target distribution one wishes to sample with an easier-to-sample probability distribution, for example at higher temperature, that we call here prior distribution.
We keep a general notation and indicate with $q(\mathbf{x}) = e^{-u_q(\mathbf{x})}/Z_q$ the prior distribution, and with $p(\mathbf{x}) = e^{-u_p(\mathbf{x})}/Z_p$ the target, where $u_p$ and $u_q$ are the respective reduced energies, $Z_q=\int e^{-u_q(\mathbf{x})} d\mathbf{x}$ and $Z_p=\int e^{-u_p(\mathbf{x})} d\mathbf{x}$ are partition functions, and $\mathbf{x}$ is the system configuration.
A set of $M+1$ replicas of the system are chosen to form a ladder of Boltzmann distributions $p_i(\mathbf{x}) \propto e^{-u_i(\mathbf{x})}$ from $p_0(\mathbf{x})=p(\mathbf{x})$ to $p_M(\mathbf{x})=q(\mathbf{x})$, such that each $p_i(\mathbf{x})$ overlaps with its neighbors in configuration space.
In the typical case of temperature expansion, one has $u_i(\mathbf{x})=(k_\mathrm{B}T_i)^{-1}U(\mathbf{x})$, where $U(\mathbf{x})$ is the potential energy, $k_\mathrm{B}$ the Boltzmann constant, and the temperatures $T_i$ interpolate between the temperatures of the target and prior distributions, $T_0=T_{\text{low}}$ and $T_M=T_{\text{high}}$, respectively.
Other kinds of expansions are also possible, such as solute tempering\cite{Wang2011} or alchemical transformations\cite{Mey2020}, generally known as Hamiltonian replica exchange.
Each replica is sampled with local moves, such as Markov chain Monte Carlo or molecular dynamics (MD), and at regular time intervals an exchange is proposed between the configurations of different replicas, $\mathbf{x}_i \rightleftarrows \mathbf{x}_j$, and accepted with probability
\begin{align}\label{E:alpha_rex}
    \alpha_{\text{REX}} =& \min \left\{1, \frac{p_j(\mathbf{x}_i)}{p_i(\mathbf{x}_i)}\frac{p_i(\mathbf{x}_j)}{p_j(\mathbf{x}_j)}  \right\} \nonumber \\ 
    =& \min \left\{1, e^{\Delta u_{ij}(\mathbf{x}_i)-\Delta u_{ij}(\mathbf{x}_j)} \right\}\, ,
\end{align}
where $\mathbf{x}_i$ and $\mathbf{x}_j$ are configurations sampled from the corresponding distribution $p_i$ or $p_j$, and $\Delta u_{ij}(\mathbf{x})=u_i(\mathbf{x})-u_j(\mathbf{x})$ is the difference in reduced energy between replica $i$ and replica $j$.
The best choice of intermediate $p_i$ is not trivial even in the well-studied case of temperature REX, and several different approaches have been proposed to optimize it\cite{Earl2005,Mey2020}.
The total number of replicas $M$ required to allow exchanges increases both with the number of degrees of freedom of the system $N$, $M\propto \sqrt{N}$, and with the distance (e.g.~in temperature) between prior and target\cite{Hukushima1996}.

One of the main limitations of REX is that a large number of replicas might be necessary to connect the target and the prior distributions, making the method computationally too expensive.
There are variants of REX that do not require a fixed number of parallel simulations, such as simulated tempering\cite{Marinari1992}, or expanded ensemble methods\cite{Lyubartsev1992, Invernizzi2020unified}, but they share the same $\sqrt{N}$ scaling of sampling efficiency as REX (see Fig.~S2 of the Supporting Information).
Several approaches have been proposed to mitigate this scaling, such as using nonequilibrium switches\cite{Ballard2009,Nilmeier2011}, but with limited practical success.
Two popular strategies to reduce the total number of replicas are to apply the tempering only to part of the system\cite{Wang2011}, or to broaden the distributions with metadynamics\cite{Deighan2012}.
We will show how to combine REX with normalizing flows to avoid altogether the need to simulate intermediate replicas.

Normalizing flows (NF) are a class of invertible deep neural networks that can be used to generate samples according to a given target distribution, and are at the core of the recently proposed Boltzmann generators\cite{Noe2019}.
A normalizing flow $f$ is an invertible function that maps a configuration $\mathbf{x}$ drawn from a prior distribution, $q(\mathbf{x})$, into a new configuration $\mathbf{x}'=f(\mathbf{x})$ that samples the output distribution of the flow, $q'(\mathbf{x})$, also called mapped distribution.
Exploiting the invertibility of $f$, one can compute the output probability density as:
\begin{equation}
    q'(\mathbf{x}')=q(\mathbf{x})\,|\det J_{f}(\mathbf{x})|^{-1}\, ,
\end{equation}
where $\det J_f(\mathbf{x})$ is the determinant of the Jacobian of $f$ and $|\det J_{f}(\mathbf{x})|^{-1}=|\det J_{f^{-1}}(\mathbf{x}')|$.
The aim of NF is to approximate the ideal map which transforms the prior into the target, thus  $q'(\mathbf{x})=p(\mathbf{x})$.
However, even without a perfect map, one can compute the importance weights $w_f(\mathbf{x})$ needed to reweight from $q'$ to $p$ the ensemble average of any observable $O(\mathbf{x})$,  $\langle O(\mathbf{x})\rangle_{p}=\langle O(f(\mathbf{x}))w_f(\mathbf{x}) \rangle_{q}/\langle w_f(\mathbf{x}) \rangle_{q}$, as discussed in Ref.~\citenum{Noe2019}.
We choose to define the weights as
\begin{equation}\label{E:weights}
    w_f(\mathbf{x}) = e^{u_q(\mathbf{x})-u_p(f(\mathbf{x}))
    +\log |\det J_{f}(\mathbf{x})| 
    }\, ,
\end{equation}
so that $w_f(\mathbf{x}) \propto p(\mathbf{x})/q'(\mathbf{x})$, where the precise proportionality constant is irrelevant for the purpose of reweighting.
In Boltzmann generators, one typically chooses as prior $q$ a normal or uniform distribution, so that independent and identically distributed samples can be used to estimate the ensemble averages $\langle \cdot \rangle_q$.
The weights are also useful to assess how effective the $f$ mapping is in increasing the overlap between $q$ and $p$, for example through the Kish effective sample size\cite{Kish1965},
\begin{equation}\label{E:neff}
    n_{\text{eff}}=\frac{\left[\sum_i^n w_f(\mathbf{x}_i)\right]^2}{\sum_i^n \left[w_f(\mathbf{x}_i)\right]^2}\, .
\end{equation}
In NF, the mapping $f$ is implemented by an invertible deep neural network that by construction has an easy-to-calculate $\det J_{f}$.
It can be trained to learn the target probability distribution $p(\mathbf{x})$ by minimizing the Kullback-Leibler divergence:
\begin{align}\label{E:kl}
    D_{\mathrm{KL}}(q' \Vert p) =& - \int q'(\mathbf{x}')\log \frac{p(\mathbf{x}')}{q'(\mathbf{x}')}\, d \mathbf{x}'
    \nonumber \\  =& - \int  q(\mathbf{x}) \log w_f(\mathbf{x})\, d \mathbf{x}- \Delta F_{pq}
    \, ,
\end{align}
where the second line has been obtained via the change of variables $\mathbf{x}'=f(\mathbf{x})$, and $\Delta F_{pq}=-\log \frac{Z_q}{Z_p}$ is the free energy difference between $q$ and $p$ that is unknown but independent of $f$.
The loss function then simplifies to $L_f = - \langle \log w_f(\mathbf{x})\rangle_q$, which in Ref.~\citenum{Noe2019} is referred to as energy-based training, as opposed to maximum-likelihood training that instead requires sampling the target distribution $p(\mathbf{x})$ and minimizing $D_{\mathrm{KL}}(p \Vert q')$.
The loss function is an upper bound to the free energy difference $L_f \geq \Delta F_{pq}$, with the identity holding only in case of the ideal map\cite{Rizzi2021}.

The main limitations of NF are that they might not be expressive enough to represent the complex map from the prior to the target, or that they can be hard to train in practice.
In particular, energy-based training is characterized by mode-seeking behavior, and struggles to converge reliably when the target distribution is multimodal\cite{Matthews2022}.

In this letter, we propose to use normalizing flows in a replica exchange setting to map the fast-mixing prior distribution to the target Boltzmann distribution, giving rise to the phase space overlap necessary for direct exchange, without the need to sample intermediate distributions.
The idea of using a map to improve the overlap between distributions was first proposed by Jarzynski under the name of \textit{targeted free energy perturbation}\cite{Jarzynski2002}, and has recently been combined with NF in the \textit{learned free energy perturbation} (LFEP) method\cite{Wirnsberger2020,Rizzi2021,Coretti2022,Falkner2022}.
By analogy, we will refer to our method as \textit{learned replica exchange} (LREX). 
Contrary to LFEP, in LREX we combine the NF mapping with local moves, which can greatly improve the sampling, as shown by several recent works\cite{Wu2020,Sbailo2021,Gabrie2021,Arbel2021,Matthews2022}.
Our approach can also be seen as a type of Boltzmann generator\cite{Noe2019}, with two main differences: (i) the prior is a nontrivial distribution sampled via MD, and (ii) the target is directly sampled in a replica exchange setting, rather than only reconstructed via reweighting.

We now present the proposed LREX method in detail.
To perform LREX, we first run a relatively short MD simulation to gather samples from the prior distribution $q(\mathbf{x})$, and use them to train the NF.
Following Ref.~\citenum{Rizzi2021}, the NF is initialized to be the identity, thus $\mathbf{x}'=\mathbf{x}$ and $q'(\mathbf{x})=q(\mathbf{x})$.
Ideally, the prior distribution should be easy to sample while still exhibiting the main features of the target, such as all the relevant metastable basins.
This choice of prior allows us to avoid most of the problems related to energy-based training of normalizing flows\cite{Noe2019,Midgley2022}, and significantly reduces computational cost.
Training in the LREX setup typically requires only a few epochs and does not suffer from mode collapse, nor from numerical instabilities linked to extremely high energies that can arise from atom clashes and other nonphysical configurations in the prior.
While the NF is training, it is possible to estimate its efficiency in increasing the phase space overlap between prior and target by computing $n_{\text{eff}}$, Eq.~\eqref{E:neff}, over the training set and/or a validation set of $n$ prior samples.
The sampling efficiency $n_{\text{eff}}/n$ also provides an estimate of the frequency with which exchanges will be accepted in the final LREX run.
Once the NF is trained, the prior and the target systems are simulated in parallel with MD and, at fixed time intervals, an exchange between mapped configurations is attempted according to the following probability:
\begin{align}\label{E:alpha_lrex}
    \alpha_{\text{LREX}} =& \min \left\{1, \frac{p(\mathbf{x}'_q)}{q'(\mathbf{x}'_q)}  \frac{q'(\mathbf{x}_p)}{p(\mathbf{x}_p)}  \right\} \nonumber \\    
    =& \min \left\{1, w_f(\mathbf{x}_q) w_{f^{-1}}(\mathbf{x}_p)  \right\} 
\end{align}
where $\mathbf{x}_p$ and $\mathbf{x}_q$ are the current configurations of the prior and target replica, respectively, and $w_{f^{-1}}$ are the weights of the inverse mapping, defined as in Eq.~\eqref{E:weights}.
If the exchange is accepted, the MD simulation of the prior continues from the configuration $\mathbf{x}_q'=f(\mathbf{x}_q)$, while the target continues from $f^{-1}(\mathbf{x}_p)$.
New velocities are randomly assigned from the respective equilibrium distributions.

\begin{figure*}
        \includegraphics[width=2\myfigwidth]{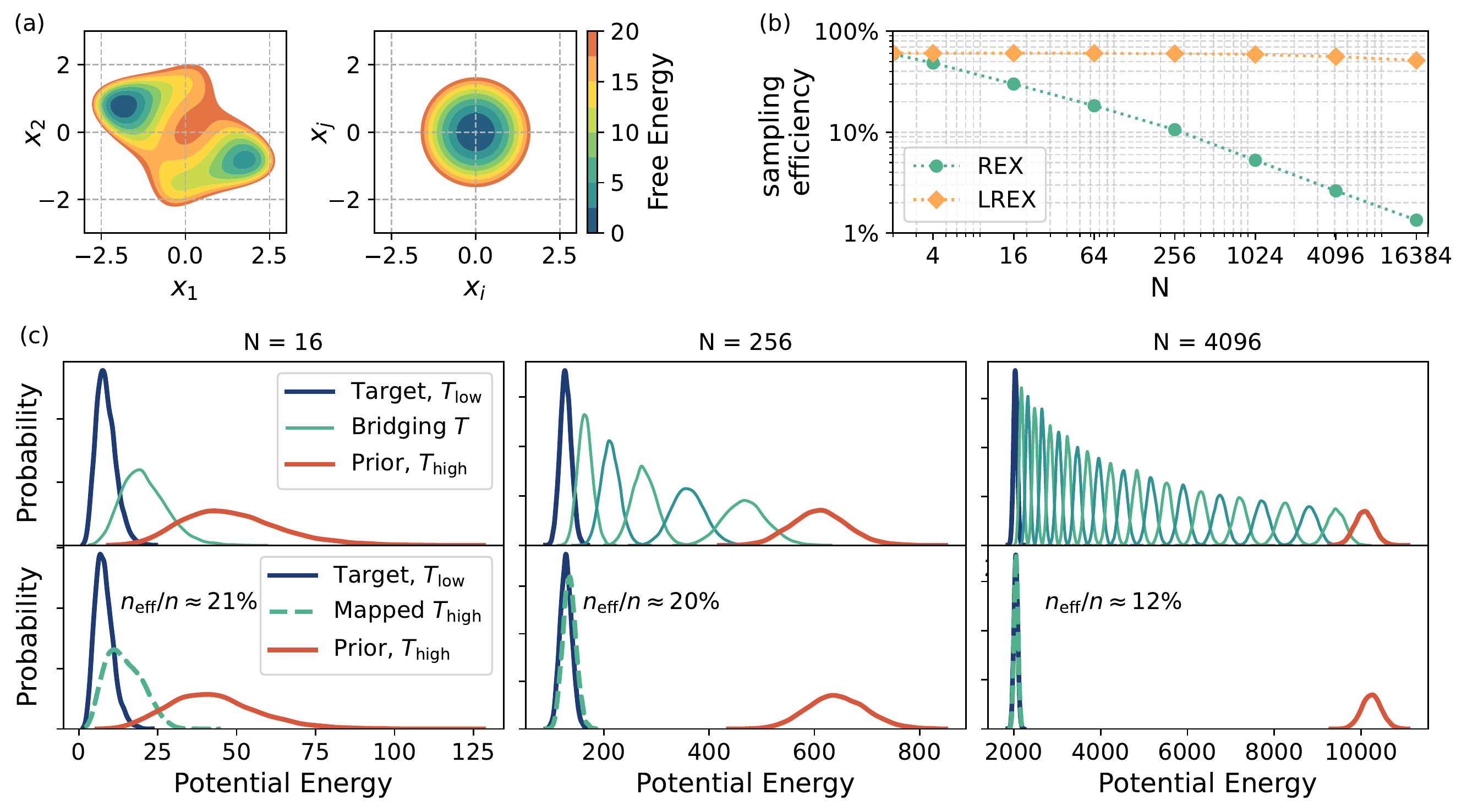}
    \caption{The $N$-dimensional double-well system. 
    (a) Apart from the first two dimensions, $x_1$ and $x_2$, all the other degrees of freedom, $x_i$, $i \in \{3, ..., N\}$, are uncoupled harmonic oscillators.
    (b) Sampling efficiency of REX and LREX, as a function of the system size $N$, in a log-log plot.
    See description in the text for details on how it is estimated.
    (c) Potential energy distribution for different dimensions $N$.
    In standard REX (top row) the number of replicas needed grows with the number of degrees of freedom, while in LREX (bottom row) the normalizing flow makes the distribution overlap, allowing for direct exchanges between prior and target at any $N$.
    The sampling efficiency, Eq.~\eqref{E:neff}, of the NF is also reported.}
    \label{F:double-well}
\end{figure*}
To demonstrate the advantages of the LREX approach over standard REX and Boltzmann generators, we present three examples.
The first system considered is a particle moving with Langevin dynamics in a $N$-dimensional double-well potential, shown in Fig.~\ref{F:double-well}a.
The first two dimensions, with coordinates $x_1$ and $x_2$, feel the 2D potential introduced in Ref.~\citenum{Invernizzi2019}, while all the other $N-2$ are subject to a harmonic potential (details in the SI).
Target and prior distribution are obtained from the same system, but at different temperatures.
The target distribution is at reduced temperature $T_{\text{low}}=1$, where transitions between the two basins are extremely rare, while the prior has $T_{\text{high}}=5$, which instead can be sampled efficiently also with a short MD run.

Figure \ref{F:double-well}c shows a visual comparison of standard REX (top row) and the proposed LREX method (bottom row).
In replica exchange, the number of intermediate temperatures grows with the system size, independently of the fact that the added degrees of freedom are trivial Gaussian noise.
Moreover, as the number of replicas increases, the mixing time required to equilibrate the simulation also increases, further limiting the efficiency of the method\cite{Abraham2008}.
In LREX, instead, only the highest and lowest temperatures need to be sampled, because the NF easily learns a transformation that makes the high-temperature configuration space overlap with the low-temperature one.
The NF architecture and training procedure are kept identical for all sizes, and are described in detail in the supporting information (SI).
Although the size of the neural network increases with $N$, the slowdown in training time and efficiency is minor, thus the overall scaling of LREX with system size is more favorable than the one of standard REX, for all the systems we studied\cite{Abbott2022}.
Figure \ref{F:double-well}b presents the sampling efficiency of REX and LREX as a function of system size $N$.
Other expanded ensemble methods would have similar scaling as REX (see Fig.~S2 in SI).
The sampling efficiency is estimated by considering the importance weights of all the replicas and is an upper bound estimate strictly valid only in the infinite simulation limit (see SI).
In a finite REX simulation, the efficiency can be further reduced by long time correlations due to a slow mixing time or small acceptance rate (see Fig.~S1 in SI).
We also do not include the training cost for LREX, which would shift the curve down by a fixed small amount, depending on the total simulation length.

\begin{figure*}
    \includegraphics[width=2\myfigwidth]{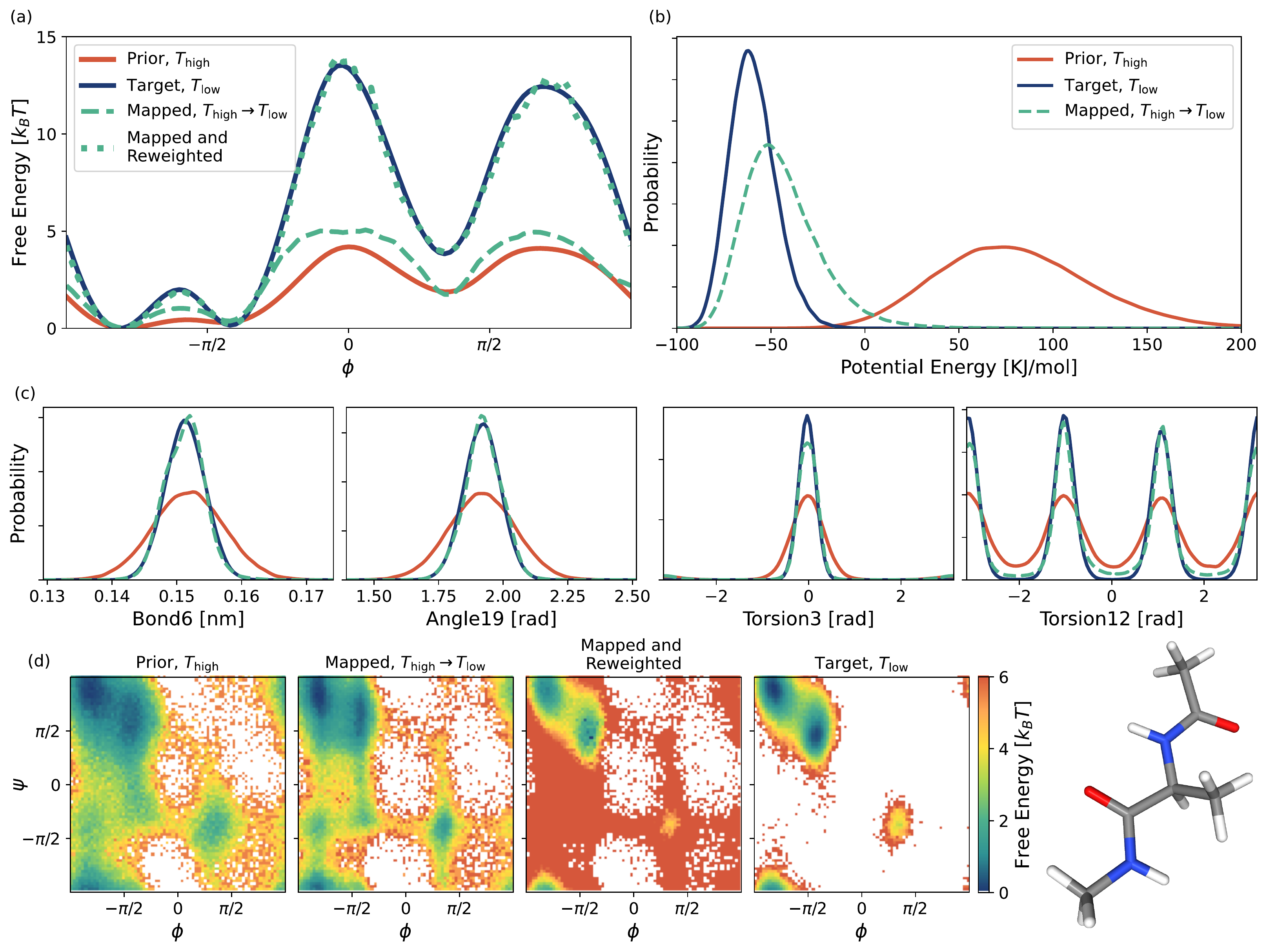}
    \caption{Alanine dipeptide in vacuum.
    (a) Free energy surface along the $\phi$ torsion angle, for the prior ($T_{\text{high}}=1000$~K), the target distribution ($T_{\text{low}}=300$~K), and the effect of the trained NF with and without reweighting.
    The NF does not move mass from one metastable state to the other, but the reweighting procedure corrects for this bias.
    (b) The energy distribution of the prior, target, and mapped distributions.
    A lack of overlap in the energy distribution implies a lack of overlap in configuration space, and without such overlap reweighting is not possible.
    (c) 1D marginal distributions over some of the internal coordinates of the molecule (48 in total).
    Notice that the NF not only has to learn the 1D marginals, but also higher dimensional correlations in the probability distribution.
    (d) Histogram of the $\phi$ and $\psi$ torsion angles sampled from the prior $q(\mathbf{x})$, the mapped $q'(\mathbf{x})$, the mapped and reweighted $w_f(\mathbf{x})q'(\mathbf{x})$, and the target distribution $p(\mathbf{x})$.
    The effect of the NF mapping is to bring configurations closer to the minima, as expected in a lower-temperature distribution.
    }
    \label{F:ala2}
\end{figure*}
Next, we consider alanine dipeptide in vacuum.
This molecule has two long-lived metastable basins that can be identified by its $\phi$ torsion angle.
The target is to sample the system at $T_{\text{low}}=300$~K, and as prior we consider a very high temperature, $T_{\text{high}}=1000$~K, at which there are no sampling issues.
With standard REX, one would need about four parallel replicas of the system to ensure proper overlap and good acceptance rate for the exchanges (see SI).
Alanine dipeptide is composed of 22 atoms, thus its configuration space has 66 dimensions, $\mathbf{x}\in \mathbb{R}^{66}$. 
Following Refs.~\citenum{Noe2019,Kohler2021} we remove global rotations and translations by modeling the NF in internal coordinates (bonds, angles, and torsion angles).
This also makes it easy to constrain hydrogen atoms at fixed bond length, and leaves us with a total of 48 degrees of freedom.
To train the NF, we perform a short 20~ns MD simulation, from which we take $20'000$ configurations at regular intervals of 500 MD steps, which are used as prior samples for the training.
Thanks to our choice of prior distributions, we can use a simple flow architecture with only a few coupling layers\cite{Durkan2019}.
Three epochs are enough to reach a good sampling efficiency $n_{\text{eff}}/n > 1 \%$, and the entire training takes only a few minutes on a single consumer-grade CPU.
Overall, the training is orders of magnitude faster than for previous Boltzmann generators of alanine dipeptide\cite{Dibak2021,Kohler2021,Midgley2022}.
The reason for this speedup is twofold.
Firstly, our prior is closer to the target, so the flow needs to learn a simpler transformation.
Secondly, the fact that LREX combines NF with local sampling of the target distribution allows one to settle for a smaller sampling efficiency, without compromising on the accuracy at which observables can be computed\cite{Gabrie2021}.

To get a sense of the effect of the trained NF map, we report in Fig.~\ref{F:ala2}c the 1D marginal distribution of some of the internal coordinates and in Fig.~\ref{F:ala2}d histograms over the 2D $\phi$ and $\psi$ plane.
Intuitively, the NF map pushes high-temperature configurations towards more low-temperature-like configurations, with smaller fluctuations from the local minima.
The NF map is far from perfect, but, as indicated by Fig.~\ref{F:ala2}b, it makes the two distributions overlap, so that reweighting can be used to reconstruct the target.
This is shown in Fig.~\ref{F:ala2}a, where the free energy surface (FES) along $\phi$ is reported.
Interestingly, the NF does not map configurations from one metastable state to the other; it mostly makes local adjustments.
This ensures that no mode collapse occurs during the NF energy-based training, which is one of the main current limitations of energy-based training for NF\cite{Ballard2009}.
In the case of alanine dipeptide, the NF we obtain is good enough that there is no need to perform the full LREX procedure; it is possible to accurately compute the torsion angle FES by direct reweighting, similarly as in Rizzi \textit{et al}.~\citenum{Rizzi2021}.
However, by running the prior and target in parallel and swapping configurations according to the LREX procedure, the sampling efficiency further increases, matching the one of standard REX for this system.
In the SI we report the case of a target distribution at a much lower temperature of $T_{\text{low}}=100$~K, where LREX is significantly more efficient than REX.

\begin{figure}
    \includegraphics[width=\myfigwidth]{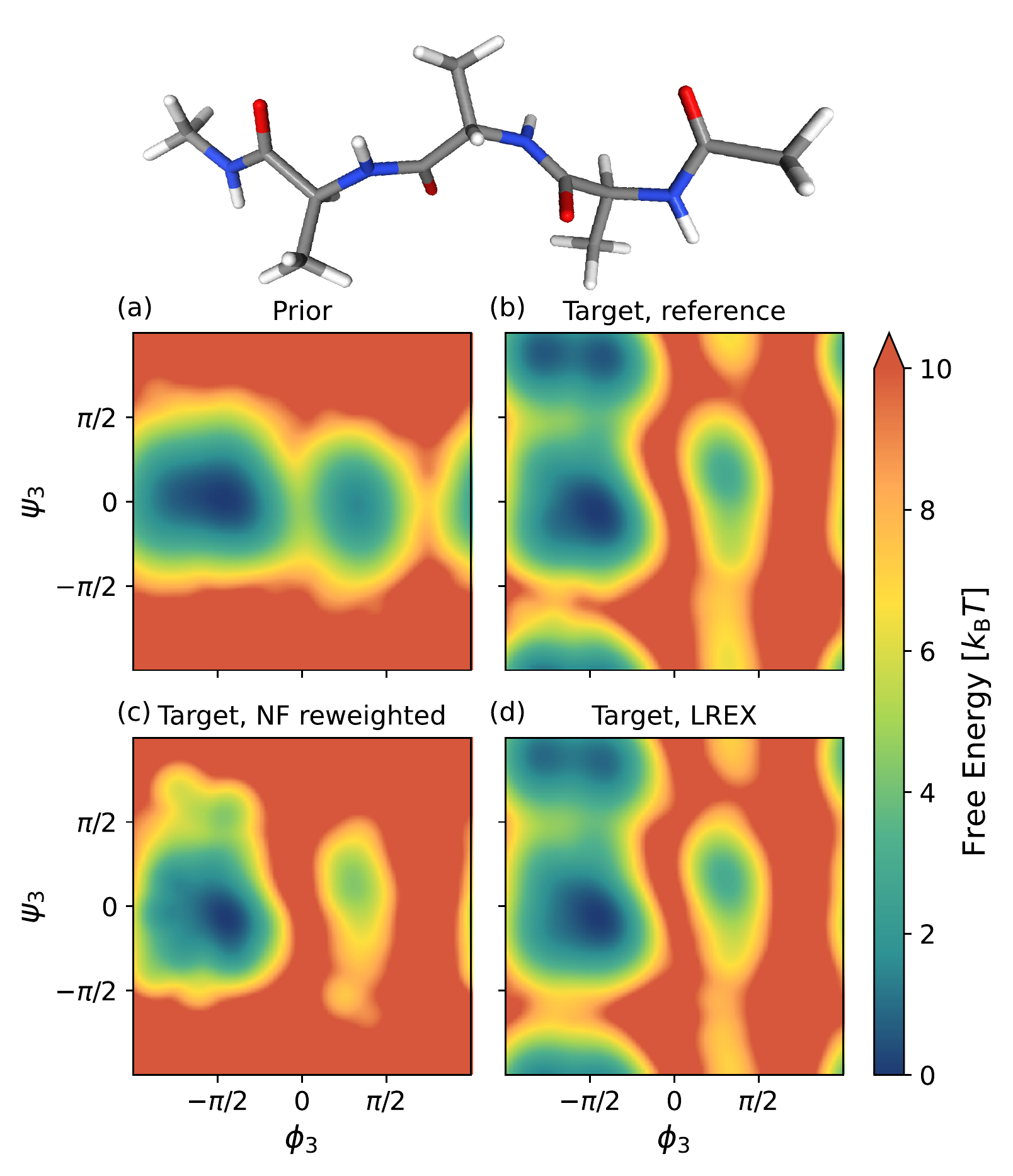}
    \caption{Free energy surfaces of alanine tetrapeptide along the $\phi_3$-$\psi_3$ torsion angles.
    (a) The prior distribution is in vacuum at $T_{\text{high}}=1000$~K and feels an extra restraint potential $U_r(\boldsymbol{\psi})$, Eq.~\eqref{E:restraint}.
    It can be efficiently sampled with plain MD.
    (b) The target distribution is in implicit solvent at $T_{\text{low}}=300$~K. The reference FES is obtained with adaptive-bias enhanced sampling on the three $\phi$ angles\cite{Invernizzi2020rethinking}.
    (c) FES obtained form the trained NF, by mapping and reweighting the prior samples.
    The NF struggles to correctly reconstruct some of the FES minima.
    However, (c) using the same NF to exchange configurations between the prior and the target in an LREX scheme, allows for an accurate estimate of the target FES.
    See Fig.~S6 in the SI for the 1D projection of these FES.
    }
    \label{F:ala4}
\end{figure}
As a last example we consider the alanine tetrapeptide molecule.
It has three pairs of $\phi$-$\psi$ torsion angles, for a total of 8 metastable basins that can be identified by the signs of the three $\phi$ angles.
The system has 42 atoms, roughly double that of alanine dipeptide, for a total of 98 internal coordinates.
This time we choose a less trivial prior and target, to show that LREX can be used not only across temperatures, but also when the potential energy between prior and target differs.
As the target we take the system in implicit solvent at $T_{\text{low}}=300$~K, while for the prior we use the system in vacuum at $T_{\text{high}}= 1000$~K and also add a restraint potential $U_r(\boldsymbol{\psi})$ acting on the three $\psi$ angles, 
\begin{equation}\label{E:restraint}
    U_r(\boldsymbol{\psi}) = \sum_{i=1}^3 k \sin^2(\psi_i/2)\, ,
\end{equation}
with $k=100$~kJ/mol.
The chosen prior distribution not only has no phase space overlap with the target, as in the previous two examples, but it also has a smaller number of local minima in some metastable basins, as can be seen in Fig.~\ref{F:ala4}a.
We use the same NF architecture as for alanine dipeptide (see SI), and the same amount of training data, obtained from just 20~ns of MD from the prior distribution.
Also in this case the training requires only a few epochs and is computationally extremely cheap.
Compared to the alanine dipeptide case, we obtain a lower sampling efficiency, $n_{\text{eff}}/n \approx 0.1 \%$, which indicates a poorer phase space overlap.
As shown in Fig.~\ref{F:ala4}c, the resulting NF fails to sample the correct target distribution when directly used via reweighting, as in Boltzmann generators and LFEP.
Using it in a LREX setup is a better strategy, and allows for an accurate reconstruction of the FES, Fig.~\ref{F:ala4}d.
To perform LREX, we concurrently run 100~ns of MD for both the prior and the target replica, attempting an exchange every 1~ps, according to Eq.~\eqref{E:alpha_lrex}.
The acceptance rate obtained is around 2\% and about $15\%$ of the accepted exchanges are useful exchanges, meaning that they make the target MD replica jump from one metastable basin into a different one.
Although it is possible to train a larger NF and use more MD data to better learn the target distribution, this would be computationally much more expensive to the extent of being less efficient than a simpler REX with many replicas.
Thanks to the LREX setup, we can make good use of a less than perfect NF, significantly improving efficiency and accuracy over the standard reweighting strategy of Boltzmann generators.

In conclusion, this letter presents the learned replica exchange method that combines replica exchange and normalizing flow, resulting in improved performance over the two methods used separately.
The LREX method has three phases: (i) run a short simulation to gather samples from the prior distribution, (ii) use these samples for energy-based training of the NF, (iii) run parallel simulations of the prior and target, attempting an exchange at regular intervals with acceptance given by Eq.~\eqref{E:alpha_lrex}.
An important aspect of the method is that it is possible to estimate the efficiency of LREX before performing the final production run, simply by computing $n_{\text{eff}}$ over the training data.
If the system is particularly challenging and $n_{\text{eff}}$ remains very low despite increasing the training or tuning the hyperparameters of the NF, it is possible to expand LREX by dividing the problem into smaller parts and training the NF to map to an intermediate system.
This provides a straightforward way to improve the accuracy of the sampling by gradually increasing the number of replicas, in analogy with REX.
It should be noted that in the worst case scenario, when the NF is completely wrong, LREX is equivalent to plain MD.

Compared to standard REX, using LREX allows for a drastic reduction of the number of replicas that have to be simulated in parallel, which can lead to a significant reduction in computation cost, especially when machine learning potentials are employed. 
This could allow the use of prior distributions that would otherwise be impractical for REX, such as coarse-grained versions of the target system\cite{Lyman2006}.
Unfortunately, a simple scaling law with system size is not available for NF, so new experiments are needed to assess feasibility on larger systems\cite{Abbott2022}.
Compared to other setups that use NF as Boltzmann generators, LREX can be significantly more efficient and handle larger systems, due to the similarity between prior and target. It can also be more accurate due to the combination with local MD sampling.
The main limitations of LREX are due to the current limitations of normalizing flows.
For example, NF have not yet been successfully applied to systems with explicit solvent, except in cases where only a subset of the system has been considered\cite{Kohler2022,Wang2022}.
However, NFs have been significantly improved in recent years, and any future improvements to the architecture or training of normalizing flows, making them more efficient or expressive, can also be readily exploited in LREX.
For this reason, we expect LREX to become a useful tool for molecular simulations, especially as normalizing flows capabilities are developing rapidly, allowing larger and more diverse systems to be handled.

\paragraph*{Data availability}
All the code and data needed to reproduce the simulations are available at \url{https://github.com/invemichele/learned-replica-exchange}.
Molecular dynamics simulations were performed with the OpenMM software\cite{openmm}.
For the reference simulations the on-the-fly probability enhanced sampling\cite{Invernizzi2020rethinking,Invernizzi2020unified,Invernizzi2022explore} has been used, as implemented in PLUMED\cite{plumed,nest}.
The normalizing flows have been implemented using the \texttt{bgflow} library (\url{https://github.com/noegroup/bgflow}).

\begin{acknowledgement}

The authors thank Jonas K\"ohler for useful discussions.
M.I. acknowledges support from the Swiss National Science Foundation through an Early Postdoc.Mobility fellowship and the Humboldt Foundation for a Postdoctoral Research Fellowship. 
F.N. and C.C. acknowledge funding from Deutsche Forschungsgemeinschaft (DFG) via CRC 1114 Project B08. 
F.N. and A.K. acknowledge funding from European Research Council (ERC) via CoG 772230 ``ScaleCell''.

\end{acknowledgement}

\begin{suppinfo}

Computational details and extra figures for the double well potential, alanine dipeptide, and alanine tetrapeptide.

\end{suppinfo}

\bibliography{refs}

\newpage\hbox{}\thispagestyle{empty}\newpage 


\section*{Supplementary material}

# Skipping the Replica Exchange Ladder with Normalizing Flows

Michele Invernizzi,*,† Andreas Krämer,†
Cecilia Clementi,‡,¶,§ and Frank Noé*,∥,†,‡,¶

†Department of Mathematics and Computer Science, Freie Universität Berlin, 14195
Berlin, Germany
‡Department of Physics, Freie Universität Berlin, 14195 Berlin, Germany
¶Department of Chemistry, Rice University, 77005 Houston, United States
§Center for Theoretical Biological Physics, Rice University, 77005 Houston, United States
∥Microsoft Research AI4Science, 10178 Berlin, Germany

E-mail: michele.invernizzi@fu-berlin.de; franknoe@microsoft.com

## Data availability

All the code and input files to reproduce the simulations presented in the paper are available at https://github.com/invemichele/learned-replica-exchange, together with the data of the main results.

## Double-well potential

We consider an $N$-dimensional double-well potential defined as

$$U_{\mathrm{dw}}(\mathbf{x}) = U_{\mathrm{wq}}(x_1, x_2) + \frac{15}{2} \sum_{i=3}^{N} x_i^2, \quad (S1)$$

where

$$U_{\mathrm{wq}}(x_1, x_2) = 2(y_1^4 + y_2^4 - 2y_1^2 - 4y_2^2 + \\ + 2y_1 y_2 + 0.8 y_1 + 0.1 y_2 + 9.28) \quad (S2)$$

is the modified Wolfe-Quapp potential introduced in Ref. 1, and $y_{[1,2]} = y_{[1,2]}(x_1, x_2)$ are rotated coordinates,

$$\begin{aligned} y_1(x_1, x_2) &= x_1 \cos(\theta) - x_2 \sin(\theta) \\ y_2(x_1, x_2) &= x_1 \sin(\theta) + x_2 \cos(\theta) \end{aligned} \quad (S3)$$

with $\theta = 0.15\pi$. The target distribution is $p(\mathbf{x}) \propto e^{-U_{\mathrm{dw}}(\mathbf{x})/T_{\mathrm{low}}}$ with $T_{\mathrm{low}} = 1$ in reduced units, while the prior one is $q(\mathbf{x}) \propto e^{-U_{\mathrm{dw}}(\mathbf{x})/T_{\mathrm{high}}}$ with $T_{\mathrm{high}} = 5$. We perform Langevin molecular dynamics (MD) with OpenMM[2] for several values of $N$, from $N = 2$ to $N = 2^{14} = 16'384$.

The replica exchange simulations are performed with geometrically spaced temperatures. The total number of replicas $M$ is chosen so that on average the acceptance rate is between 15% and 20%. Exchanges are attempted every 100 MD steps, and we run each replica for a total of $1'000'000$ steps. As expected, a significant increase of the mixing time is observed as $N$ and $M$ increase. This can be seen e.g. from the trajectory of the lower temperature replica, Fig. S1a, where the number of useful exchanges between different metastable basins is significantly smaller for higher $N$.

The normalizing flow (NF) used is an affine flow with two coupling layers obtained by splitting in half the system's dimensions. The neural networks used have 4 layers of size $[N/2, 128, 128, N/2]$, for each of the studied system sizes $N$. The NF is initialized to the identity by setting all parameters to zero. For the training we collect $64'000$ samples from the prior, one each 100 MD steps. The size of training data has not been optimized, and a smaller

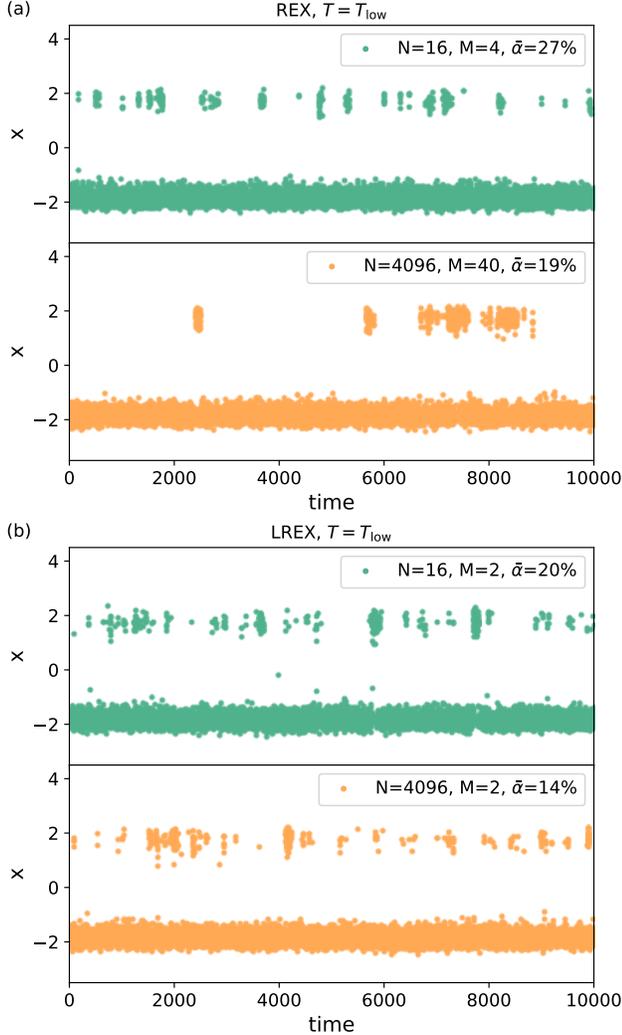

Figure S1: The $x$ coordinate trajectories of the low temperature replica of the double-well system obtained via standard REX (a) and LREX (b). For each of the trajectories we report the system size $N$, the total number of replicas $M$, and the average acceptance rate $\bar{\alpha}$.

one would likely work just as well. Training is performed with the Adam optimizer for one epoch with batch size 128 and the learning rate is progressively reduced with a scheduler from $10^{-2}$ to $10^{-5}$. We found the learning rate to be an important hyperparameter. We hypothesise that since the NF is initialized to the identity, a large learning rate is needed to move it out of this low entropy state, while in order to precisely converge to the correct high-dimensional target, a very small learning rate is essential to avoid noise that can kill the sampling efficiency. Once the NF is trained, LREX can be run by simulating only the high and low temperature and directly exchanging configurations via the NF. The observed average exchange rate is well approximated by the Kish effective sampling size measured on the training data using the NF weights. It is around 20% for all $N$ values, except for the two largest tested systems, $N = 2^{12}$ and $N = 2^{14}$, for which it lowers to about 17% and 12% respectively, as reported in Fig.1b of the main text. Contrary to REX, in LREX we do not observe mixing issues, since the exchanges are only between two replicas and the acceptance rate is more than large enough for any size $N$, Fig. S1b.

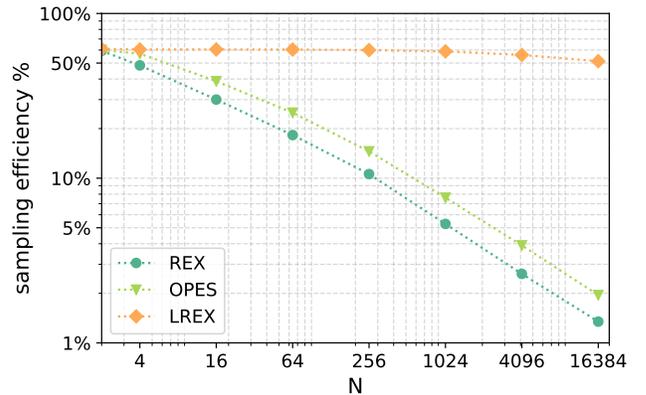

Figure S2: Sampling efficiency of the double-well system as a function of the size $N$ for different methods. Data for REX and LREX are the same shown in Fig. 1b of the main text. OPES samples the same ensemble as REX, but with a single replica, thanks to an adaptive bias potential.[3] Despite being more efficient than REX, OPES follows the same scaling law.

Figure S2 reproduces Fig. 1b of the main text,

with the addition of the sampling efficiency obtained with the on-the-fly enhanced sampling method (OPES).[3] OPES can be used to sample an ensemble equivalent to the one sampled by REX, but in a single MD run, by adding an adaptive bias potential. As can be seen from Fig. S2, OPES is more efficient than REX, but presents the same efficiency scaling as a function of system size, $\eta \propto \sqrt{N}$. The number of samples needed to converge OPES adaptive potential increases with system size, but the resulting slow down is less dramatic than the one due to slow mixing in REX.

The sampling efficiency $\eta$ shown in Fig. S2 and Fig. 1b of the main text is calculated as follows:

$$\eta = \frac{1}{Mn} \sum_i^M n_{\text{eff}}^{(i)}, \qquad (S4)$$

where $M$ is the total number of replicas, $n$ is the total number of samples and $n_{\text{eff}}^{(i)}$ is the effective sampling size of replica $i$, calculated from the importance weights according to Eq. (4) of the main text. For the LREX case these are the weights given by the NF, Eq. (3) main text, while for REX they are calculated as

$$w_i(\mathbf{x}) = e^{(T_i^{-1} - T_{\text{low}}^{-1})k_B^{-1}U(\mathbf{x})}, \qquad (S5)$$

where $T_i$ is the temperature of replica $i$. Thus the $n_{\text{eff}}$ of the low temperature replica is always equal to $n$, as expected, and in REX only the nearest replica usually has a $n_{\text{eff}} > 1$. From a practical point of view, unless $M$ is very small, one has that $\eta_{\text{REX}} \approx \frac{1}{M}$. Instead for LREX the contribution to the sampling efficiency coming from the high temperature replica remains higher than 2.5% even for $N = 2^{14}$, thus $\eta_{\text{LREX}} > 50\%$ for each $N$, as shown in Fig. S2. Finally, in the case of OPES where only one replica of the system is simulated, the weights to calculate $n_{\text{eff}}$ are computed according to the value of the bias potential, as explained in Ref. 3.

As a final remark, we notice that for this simple double-well system it is also possible to analytically define a simple mapping $f^*$ from high temperature to low temperature that allows for direct exchanges. This mapping does not change the first two nontrivial coordinates and properly rescales the remaining ones:

$$f^*(x_i) = \begin{cases} x_i & \text{if } i \in \{1,2\} \\ x_i \sqrt{\frac{T_{\text{low}}}{T_{\text{high}}}} & \text{if } i \in \{3,N\} \end{cases} \qquad (S6)$$

For $N = 2$ the $f^*$ map is the identity, and its efficiency is $n_{\text{eff}}/n = 17\%$, lower than what the trained NF was able to achieve, that is around 22%. Contrary to the NF, $f^*$ exactly maps the added degrees of freedom and as a consequence its efficiency does not depend on $N$, meaning that it remains constant at 17% no matter how much we increase the system size. The catch it that such an analytical map is extremely difficult to guess for any system but the most trivial ones. Already for alanine dipeptide it would be quite difficult to find one, while the NF can easily be trained to learn it.

## Alanine dipeptide

The alanine dipeptide molecular dynamics simulations where performed with OpenMM[2] using a Langevin integrator with standard parameters (see GitHub repository). To obtain a reference we run a multithermal simulation using the on-the-fly probability enhanced sampling method (OPES).[3] OPES provides a simple way to sample all the relevant temperatures in a single MD simulation, by adding an adaptive bias. The ensemble sampled at convergence is equivalent to the one sampled by REX, and the sampling efficiency has the same scaling, albeit being typically higher.

The normalizing flow is implemented using the bgflow library (https://github.com/noegroup/bgflow), and acts on the internal coordinates of bonds, angles and torsions. It consists of two spline coupling layers[4] that act separately on bonds, angles and torsions. Circular splines are used for torsion angles, to handle the periodicity. For bonds and angles, the domain of the spline flows are defined using the high temperature training data, with the assumption that for the target low temperature it will be the same or smaller. The NF is initialized to the identity by setting all 438′842

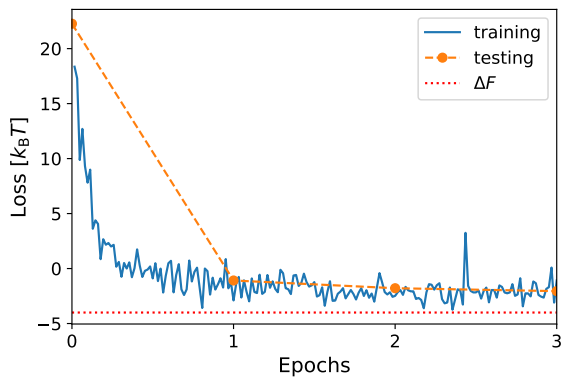

Figure S3: Typical training of the alanine dipeptide normalizing flow. The loss function is an upper bound to the free energy difference $\Delta F$ between the two temperatures, $T_{\text{high}} = 1000$ K and $T_{\text{low}} = 300$ K. The testing data is taken from another MD run at $T_{\text{high}}$.

parameters to zero. Training by energy is done with the Adam optimizer and a fixed learning rate of $10^{-3}$ and a small batch size of 32. We already find good results after 3 epochs over the $20'000$ samples from the prior (20 ns), see Fig. S3. Similar results are obtained also when using half the training points, a 10 ns MD run. We clip contributions to the energy that are too high, to make training more stable. Once the flow is trained, we run LREX for 100 ns, simulating both prior and target while trying an exchange every 1 ps (500 MD steps). The average acceptance rate is approximately 23% and among the accepted exchanges approximately 4% move the low temperature replica from one metastable state to the other one.

With the same setup we can train the NF using as target $T_{\text{low}} = 100$ K instead of 300 K. The resulting sampling efficiency is smaller, $n_{\text{eff}}/n \approx 0.1\%$, but still high enough to be useful. In this case the flow is not good enough to reconstruct the FES along the $\phi$ angle from reweighting alone, so we must use LREX to have an accurate estimate. Given the extremely low temperature of the target, the free energy difference between the two metastable states is very large, $\approx 10 k_{\text{B}} T$. As a consequence, we need a much longer simulation to obtain a relevant amount of samples also from the least populated metastable state. We report the results in Fig. S4.

## Alanine tetrapeptide

The reference simulations for alanine tetrapeptide were obtained with OPES[5] by enhancing the sampling of the three $\phi$ angles. The very same NF setup used for alanine dipeptide is also used for alanine tetrapeptide, with no fine tuning. The resulting flow is slightly bigger, with $792'022$ parameters. The sampling efficiency $n_{\text{eff}}/n$ of the trained NF is approximately 0.8%, and the acceptance rate of the LREX exchanges over the 100 ns MD run is approximately 2.4%. Of the accepted exchanges about 12.6% make the target replica move to a different metastable state. Figure S5 shows the two dimensional FES estimates as Fig. 3 from the main text, but for all Ramachandran angles, while Fig. S6 reports the one dimensional FES for the same data.

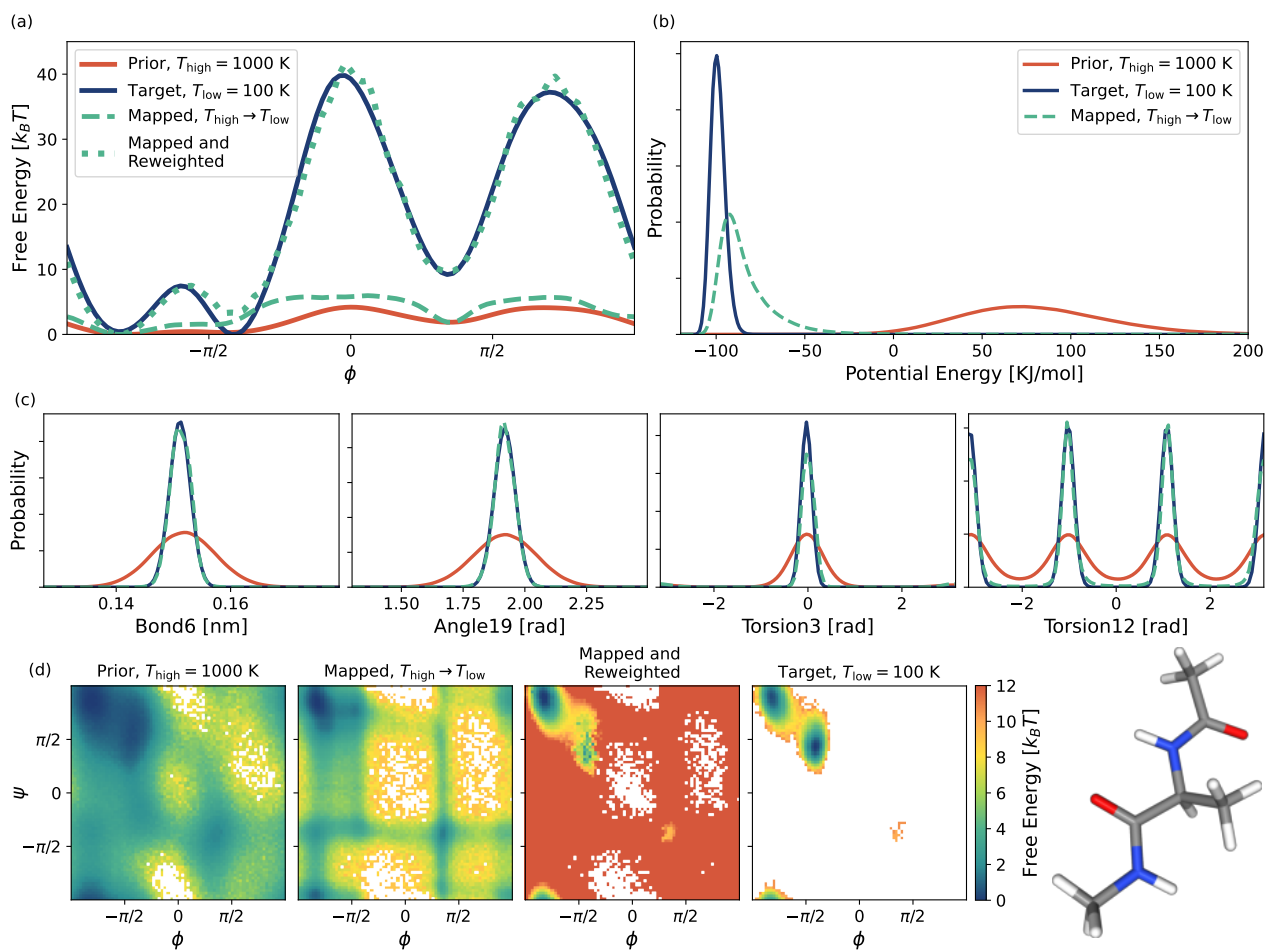

Figure S4: Alanine dipeptide results, same as Fig. 2 of the main text, but with a different target distribution at $T_{\text{low}} = 100$ K. Notice the different range of the y axis in (a) and of the color bar in (d) - the prior looks quite different but is actually the same. From (b) one can see that the prior and the target are much more distant than for the case reported in the main text, Fig. 2b. To perform REX one would thus need more than double the computational cost, while for LREX it remains roughly constant. The histograms reported in (d) are obtained from a 1 $\mu$s LREX run (10 times longer than the one from Fig. 2 of the main text).

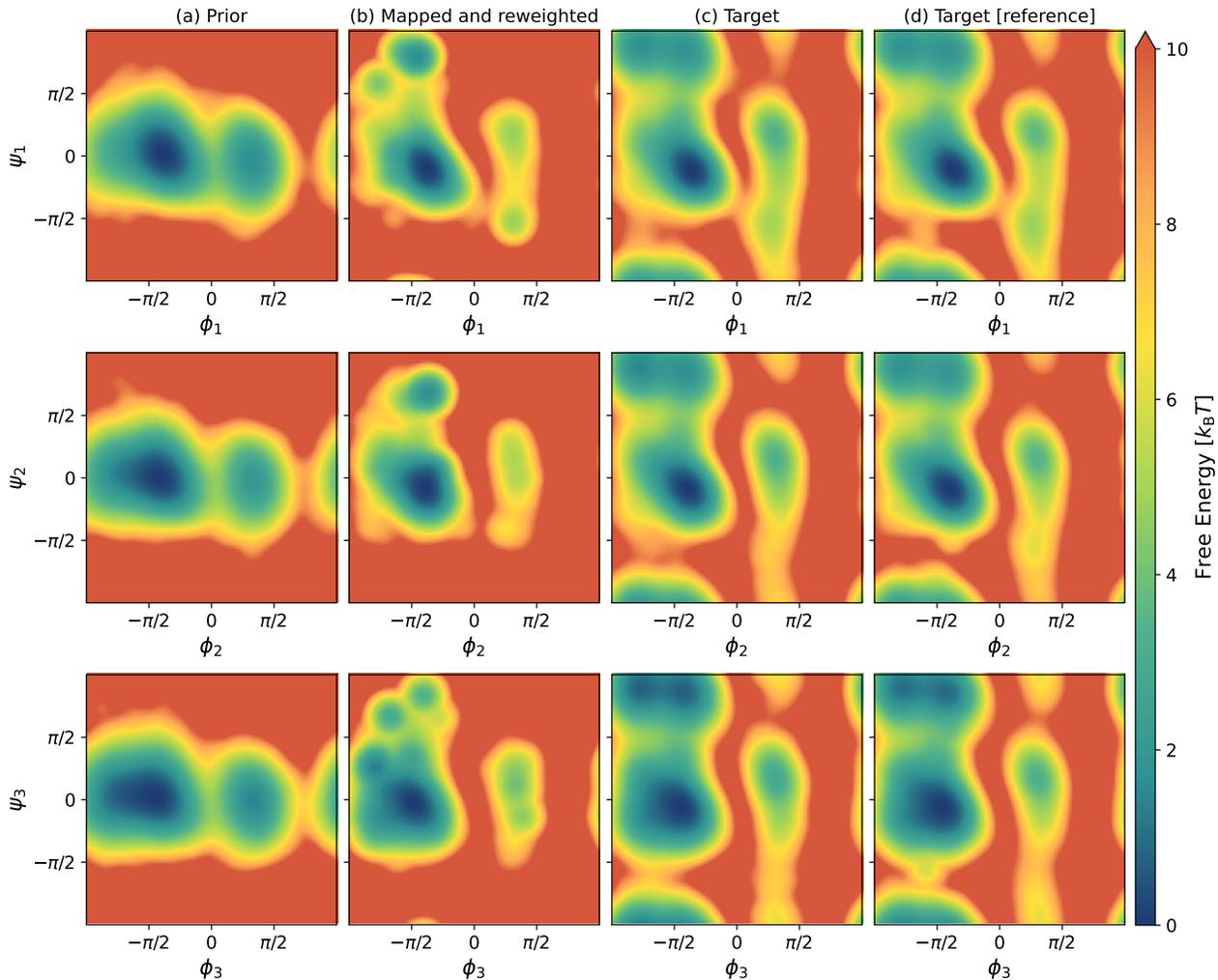

Figure S5: Free energy surfaces of alanine tetrapeptide along the three $\phi$-$\psi$ torsion angles, obtained from an LREX run. (a) The prior distribution is in vacuum at $T_{\text{high}} = 1000$ K and feels an extra restraint potential $U_r(\boldsymbol{\psi})$, Eq. (7) of the main text. It can be efficiently sampled with plain MD. (b) Shows the FES estimate we obtain form the trained NF, by mapping and reweighting the prior samples. The NF struggles to correctly reconstruct some of the FES minima. However, (c) we use this same NF to run exchange configurations between the prior and the target in an LREX scheme, and this results in an accurate estimate of the target FES. (d) For reference, we report the target FES obtained with adaptive-bias enhanced sampling on the $\phi$ angles.[5]

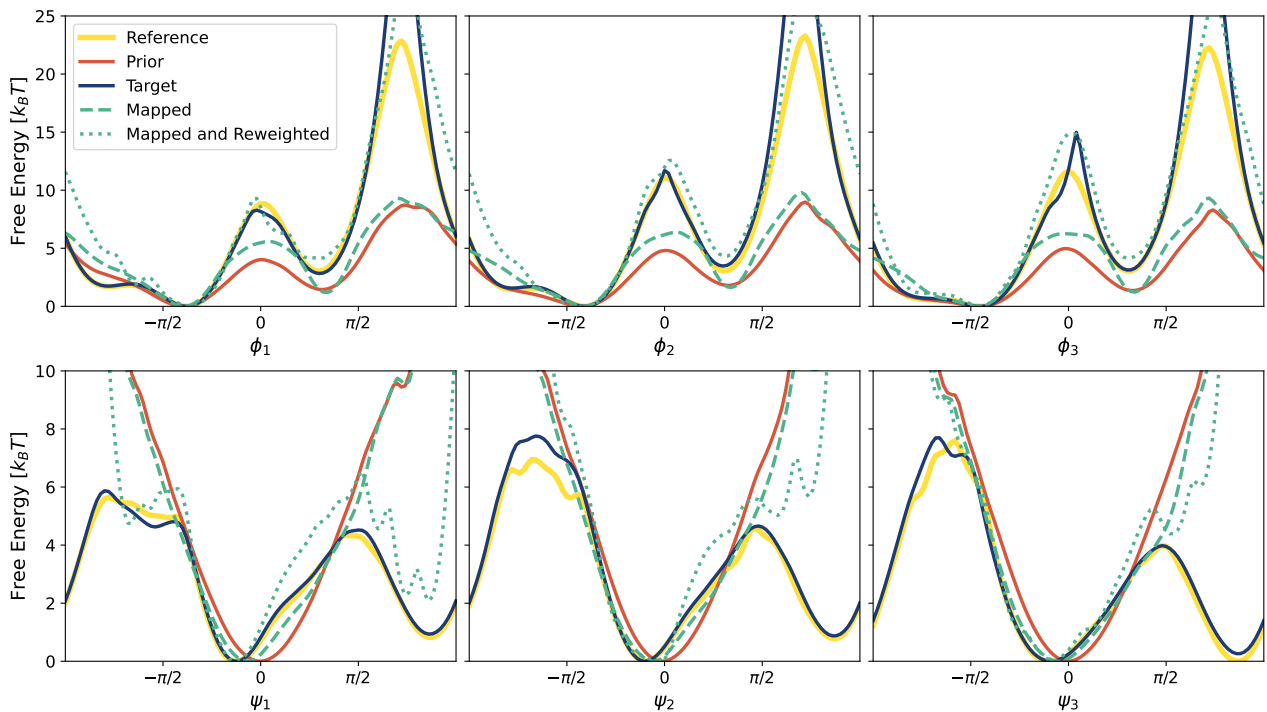

Figure S6: The same alanine tetrapeptide data used for Fig. S5 and Fig. 3 of the main text here is used for reconstructing the 1D FES along the three $\phi$ and $\psi$ torsion angles. As before, the prior and the target data come from a LREX run, while the reference is from an OPES run, both 100 ns long. Notice how the prior along the $\psi$ angles is completely reshaped by the extra restraint potential $U_r(\boldsymbol{\psi})$, and the trained NF is not powerful enough to recover the correct profile, even after reweighting. The target sampled with LREX instead is in very good agreement with the reference, except for very high free energy regions, that do not contribute to ensemble averages.



\end{document}